\pdfoutput=1 
\documentclass[nofootinbib,prl,aps,reprint,amsmath,amssymb]{revtex4-1}
\usepackage{hyperref}
\usepackage{changes}
\usepackage[section]{placeins}
\usepackage[titletoc]{appendix}
\usepackage{xcolor}
\usepackage{dsfont}
\usepackage{bm}
\usepackage{mathtools}
\usepackage{enumerate}
\DeclareMathAlphabet{\mathpzc}{OT1}{pzc}{m}{it}

\newcommand{\sayy}[1]{`#1'}
\DeclarePairedDelimiter\abs{\lvert}{\rvert}%
\usepackage{calc}
\usepackage{accents}

\providecommand{\href}[2]{#2}

\usepackage{amsthm}

\def\be{\begin{equation}}
\def\ee{\end{equation}}
\def\bea{\begin{eqnarray}}
\def\eea{\end{eqnarray}}

\def\sig{\sigma}

\def\la{\langle}
\def\ra{\rangle}
\def\Eu{ \mathfrak{H} }

\def\obs{\mathcal{O}}
\def\emi{\mathcal{E}}

\definecolor{MyB}{rgb}{0.1,0.1,1.0}

\def\apj{{Astrophys. J.} }

\def\jcap{{J. Cosmol. Astropart. Phys.} }

\def\prd{{Phys. Rev.} D }

\def\apj{{Astrophys.\ J.}}
 
\def\apjl{{Astrophys. J. Lett.}}

\def\MNRAS{{Mon.\ Not.\ R.\ Astr.\ Soc.}}

\def\prd{{{Phys.\ Rev. D.}}}

\begin{document}
\title{Bulk flows in the local Universe and the importance of relativistic effects} 
\author{Asta~Heinesen}
\email{asta.heinesen@nbi.ku.dk}
\affiliation{Univ Lyon, Ens de Lyon, Univ Lyon1, CNRS, Centre de Recherche Astrophysique de Lyon UMR5574, F--69007, Lyon, France} 
\affiliation{{Niels Bohr Institute, Blegdamsvej 17, DK-2100 Copenhagen, Denmark}}

\begin{abstract} 
Bulk flow velocities are typically estimated in the idealised picture where observers are moving within a perfectly homogeneous and isotropic space-time. 
This picture is consistent within standard perturbation theory up to relativistic effects that lead to correction terms of order $v z$, where $z$ is the redshift of observation, and $v$ is the amplitude of the bulk flow. 
The dominant relativistic contributions at scales $z \lesssim 1$ are caused by gravitational redshift and time evolution of the velocity field. 
We include these effects within a broadly applicable weak-field approximation, and provide a cosmographic formula for estimating bulk flows at a high precision.   
Based on this formula, we judge that recent bulk flow estimates are biased toward larger values by $\sim 10\%$. 
This theoretical bias surpasses the measurement biases of the same estimates, and it will become still more important to account for the relativistic effects as the scales at which bulk flows are estimated to increase.

\end{abstract}

\maketitle

\section{Introduction}
\label{sec:intro} 
Large-scale matter flows in our cosmic neighbourhood have been a subject of investigation for decades \cite{1982ApJ...258...64A,1987ApJ...313L..37D,1993ApJ...412....1D,1994ApJ...425..418L,2016IAUS..308..336M,Hoffman:2015waa,Scrimgeour:2015khj}. 
These cosmic matter flows, also denoted bulk velocities, serve as important probes of the large-scale mass density and gravitational theory \cite{1987Natur.327..210P,1991Natur.349..753K}.  
In recent years, several analyses have identified coherent flows of matter with a bigger amplitude than expected within the Lambda cold dark matter ($\Lambda$CDM) model at scales $\gtrsim 100$ Mpc/$h$ \cite{Kashlinsky:2009dw,2010MNRAS.407.2328F,2013MNRAS.430.1617L,2016IAUS..308..336M,Peery:2018wie,Watkins:2023rll}. 
Some analyses indicate the larger-than-expected bulk flows for distance scales $\gtrsim 300$Mpc$/h$ \cite{Kashlinsky:2008ut,Kashlinsky:2009dw} while others show consistency with the $\Lambda$CDM expectation or remain inconclusive \cite{Feindt:2013pma,Planck:2013rgv}. There are forecasts preparing for improved high-redshift constraints of the bulk velocity field \cite{Nusser:2011sd,Mathews:2014fma}. 
Some of the most recent analyses find a growing trend in the amplitude of matter flows over the distance range from $100$ to $300$Mpc$/h$ \cite{Peery:2018wie,Watkins:2023rll}. This is unexplained within the $\Lambda$CDM model where a decay of the bulk velocity amplitude $\propto 1/(\text{distance})$ is generically expected within the standard inflationary scenario \cite{Kashlinsky:2009dw}. 

The targeted astrophysical sources are typically modelled as having peculiar-motion relative to the canonical frame in a Friedmann-Lema\^itre-Robertson-Walker (FLRW) universe model; see, e.g., \cite{Davis:2014jwa}. 
This is a self-consistent lowest-order approximation of the velocity field within FLRW perturbation theory when the scale of observation remains much smaller than the Hubble horizon.   
When the scale of observation is a significant fraction of the Hubble length scale, gravitational redshift must be accounted for in a precise treatment of bulk flows \cite{Pyne:2003bn,Hui:2005nm}. In addition, since the velocity field is not observed at an instance of cosmic time (speed of light is finite), one must generally account for the systematic evolution of the velocity field with time \cite{Kaiser:2014jca}. 
These relativistic imprints may be of particular importance when the signal that is to be determined, such as bulk velocity, is itself small. 
If not included, they can lead to theoretical biases. 
These are distinct from measurement biases, caused by, for instance, non-Gaussian distributed errors for the measured distances \cite{Watkins:2014oia} and sampling effects \cite{Andersen:2016ywu}.  

Relativistic effects are important in the vicinity of very dense objects, but also toward large scales; see table~1 in~\cite{Buchert:2009wj} for estimates of gravitational potentials for various systems. 
Surveys have now reached the depth in redshift and measurement accuracy that makes it important to include the above-mentioned relativistic effects. 

In this paper, we consider a generic weak-field setting within which the distortion of the distance--redshift relation coming from motion effects and intrinsic impacts of the photons (gravitational redshift, the integrated Sachs--Wolfe effect, and lensing) are described covariantly, {and we provide a formula for accurate estimates of bulk flows within this broad setting.} 

{The outline of this paper is the following: We first describe the weak-field setting and matter model that we assume. We then describe the distances and redshifts to astrophysical sources within this setting, and we}
 perform a cosmographic series expansion, which allows one to evaluate the peculiar Hubble law for luminosity distance consistently while including the above-mentioned relativistic effects.  
We {finally} discuss the correction terms that arise and their impact on bulk flow measurements {in the literature}.

\vspace{3pt} 
\noindent
\underbar{Notation and conventions:}
We use units where $c=1$ and $G=1$, where $c$ is the speed of light and $G$ is the gravitational constant. Greek letters $\mu, \nu, \ldots$ are used for space-time
indices in a general coordinate system. 
The signature of the space-time metric $g_{\mu \nu}$ is $(- + + +)$ and the connection $\nabla_\mu$ is the Levi-Civita connection.

\section{Weak-field approximation} 
In our analyses of bulk motions, we assume the following: 
(i) There exists a \emph{quasi-Newtonian} frame that is both irrotational and shear free. This may be seen as neglecting gravitational waves and relativistic frame dragging; 
(ii) The matter frame (which may have both vorticity and shear) is a dust source, i.e. pressure-free matter, in the presence of a cosmological constant; and (iii) The matter has a velocity field relative to the quasi-Newtonian frame that is {everywhere much smaller than the speed of light.} 

This approximation largely follows that in \cite{vanElst:1998kb}, but here we allow for a nonzero cosmological constant. 
The approximation includes, but is not limited to, the linearly perturbed FLRW models with ordinary matter and a gravitational constant.  
For details on the results reviewed in this section, we refer the reader to \cite{1996PhDT........25V,vanElst:1998kb}.

\subsection{The quasi-Newtonian space-time}
\label{sec:QN} 
The quasi-Newtonian space-times that we will consider are characterised by having a stable frame that is expanding isotropically, and which therefore bears resemblance to a canonical FLRW frame in terms of its kinematical properties. 
Mathematically, we require the existence of a time-like congruence with 4-velocity field $n^\mu$ with kinematical decomposition 
\bea
\label{def:nablan}
\nabla_{\mu}n_\nu  &&= \frac{1}{3}\theta h_{\mu \nu }    +    \sig_{\mu \nu}  +  \omega_{\mu \nu}  - n_\mu a_\nu \, , \\  %= \frac{1}{3}\theta h_{\mu \nu } - n_\nu a_\mu  \ , \nonumber \\ 
 a_{\mu} && \equiv n^\nu \nabla_\nu n_\mu  \,   ;  \qquad  \theta \equiv \nabla_{\mu}n^{\mu}  \, ;  \nonumber  \\  
 \sig_{\mu \nu}  && \equiv h_{  \la \mu  }^{\, \alpha}  h_{ \nu \ra }^{\, \beta }\nabla_{  \alpha}n_{\beta  }  \,  ; \qquad  \omega_{\mu \nu} \equiv h_{  [\mu  }^{\, \alpha}  h_{  \nu] }^{\, \beta }\nabla_{  \alpha}n_{\beta  }    \,  ,   \nonumber
\eea 
{where $\theta$ is the isotropic volume expansion rate, $\sig_{\mu \nu}$ is the anisotropic deformation (shear), and $\omega_{\mu \nu}$ is the rotation (vorticity) of the congruence,} 
such that shear and vorticity vanish{:}
\bea
\label{def:zeroshearvorticity}
\sig_{\mu \nu} = 0 \, ,  \qquad \omega_{\mu \nu} = 0 \, . 
\eea 
{In such a frame there can be no gravitational wave distortion or frame dragging effects that are associated with anisotropic distortions and/or rotational degrees of freedom of space, and thus, due to the absence of these relativistic effects, we shall refer to $n^\mu$ as a quasi-Newtonian 4-velocity field. 
Mathematically, we have that the vanishing shear condition implies that the lie transport of the spatial metric tensor $h_{ \mu \nu }$ along the flow lines of $n^\mu$ is given by a conformal scaling: 
\bea
\label{def:lie}
&&  \pounds_{\bm n} h_{\mu \nu } \equiv  2 h^{\alpha}_{\nu} h^{\beta}_{\mu} \nabla_{(\alpha} n_{\beta)}  =  \frac{2}{3}\theta h_{\mu \nu }  \, . 
\eea  
Thus, the flow lines of $n^\mu$ are parting isotropically with time, much as in an FLRW universe model, except that $\theta$ may be a complicated inhomogeneous function and thus generally does \emph{not} reduce to $3H$, where $H$ is the homogeneous Hubble parameter in FLRW models. } 
In the above,  
we have used the projection tensor onto the three-dimensional space orthogonal to $n^\mu$: 
\bea
\label{prroj} 
h_{ \mu  \nu } \equiv n_{ \mu } n_{\nu } + g_{ \mu  \nu }  \, , 
\eea 
and the notation with $[]$ around indices indicating anti-symmetrisation in the involved indices, and $\la \ra$ singling out the symmetric and trace-free part of the tensor with respect to the spatial metric \eqref{prroj}.  
Vanishing of vorticity implies that $n_\mu$ can be written as
\bea
\label{gradient} 
\hspace*{-0.6cm}  n_\mu  = - N \nabla_\mu t  \, , \quad \;  N \equiv \frac{1}{n^\mu \nabla_\mu t} = \frac{1}{ \sqrt{ - g^{\mu \nu} \nabla_\mu t \nabla_\nu t }}  \, , 
\eea 
where the time-lapse, $N$, ensures the appropriate normalisation $n^\mu n_\mu = -1$. 
Here, $t$ is a time parameter that labels spatial hypersurfaces in the quasi-Newtonian frame. 
It follows that the 4-acceleration of the quasi-Newtonian frame is irrotational: 
\bea
\label{vorticitydot}
\hspace*{-0.8cm}  
a_\mu = D_\mu \ln(N) \, ,   
\eea 
where $D_\mu$ is the spatial covariant derivative, defined through its acting on a tensor field: $D_{\mu} T_{\nu_1  \nu_2  \dots }^{ \quad \quad \gamma_1  \gamma_2 \dots  } \equiv  h_{ \nu_1 }^{\, \alpha_1 } h_{ \nu_2 }^{\, \alpha_2 } \dots     \,  h_{ \beta_1 }^{\, \gamma_1 } h_{ \beta_2 }^{\, \gamma_2 } \dots    \, h_{ \mu }^{\, \sigma } \nabla_\sigma  T_{\alpha_1  \alpha_2 \dots  }^{ \quad \quad \; \beta_1  \beta_2 \dots }$.

\vspace{6pt}
\noindent 
\noindent\fbox{%
    \parbox{0.46\textwidth}{%
\noindent 
\textbf{Quasi-Newtonian 4-acceleration.}  
The field $a^\mu$ has interpretation as the external force per unit mass that must be applied to counteract any gravity from structures that would anisotropically distort the pattern of geodesic particles and cause shearing of their congruence. It is analogous to \emph{minus} the Newtonian 3-acceleration of a freely falling test particle. The lapse function $N$ has interpretation as the gravitational time dilation, and its logarithm is analogous to the Newtonian gravitational potential. 
    }
}

\vspace{6pt}
\noindent 
It follows from \eqref{def:zeroshearvorticity} that the magnetic part of the Weyl tensor in the quasi-Newtonian frame is identically zero \cite{1996PhDT........25V}: 
\bea
\label{magnetic} 
H_{\alpha\beta} \equiv
-\frac{1}{2}\epsilon_{\rho\sigma\gamma\delta}
C_{\mu\nu}{}^{\gamma\delta}n^{\rho}h_{\alpha}{}^{\sigma}
n^{\mu}h_{\beta}{}^{\nu}  =  0  \, . 
\eea

\subsection{The matter and its peculiar-motion}
\label{sec:dust} 
We assume that the Universe is well described by a dust source at the {observational scales over which one aims to measure the bulk flow; i.e., we neglect any effective pressure, anisotropic stress, and heat flux terms that may arise from \emph{internal} motions within the effective cosmic fluid parcels defined at these scales.} We thus require that the energy momentum tensor takes the form  
\bea
\label{T}
T_{ \mu \nu } =  \rho \, u_{ \mu } u_{\nu } \, , 
\eea  
where $\rho$ is the restmass density and $u^\mu$ is the 4-velocity of the matter frame. Under this assumption, Einstein's equations including a cosmological constant, $\Lambda$, read 
\bea
\label{TG}
R_{ \mu  \nu } - \frac{1}{2} R g_{ \mu \nu } + \Lambda g_{ \mu \nu } = 8 \pi G  T_{ \mu \nu }\, , 
\eea  
where $R_{ \mu  \nu }$ is the Ricci tensor and $R$ is the Ricci scalar of the space-time.

It follows directly from the dust form of the energy momentum tensor \eqref{T} and its conservation equation $\nabla_\nu T^{ \mu \nu } = 0$ that the matter source is geodesic: $u^\nu \nabla_\nu u^\mu = 0$. 
We can perform the following general decomposition of the matter 4-velocity: 
\bea
\label{udecomp}
 u^\mu = \gamma (n^\mu + v^\mu) \, , \quad \gamma \equiv - n^\mu u_\mu  \,  , \quad n^\mu v_\mu = 0 \, , 
\eea 
where we denote $v^\mu$ the \emph{peculiar-velocity field} of the matter relative to the quasi-Newtonian rest frame, and where we can express the Lorentz factor in its usual form $\gamma = (1 - v^\mu v_\mu )^{- \frac{1}{2}}$. 
{Generally the matter frame will exhibit shear, vorticity, and magnetic Weyl curvature that may be quantified entirely in terms of $v^\mu$ and its gradients; cf. \cite{vanElst:1998kb}. 
Thus, the principle of relativity, that the kinematics of matter is intimately related to the structure of space-time through curvature is directly reflected in these relations.  } %peculiar-velocity field is intimately related to the curvature of the space-time. This is one manifestation of the principle of general relativity, that  curvature:  
%We {This velocity field can be viewed as a large-scale peculiar }
%{This peculiar-velocity field is more generally-defined than in the $\Lambda$CDM scenario with a dynamical dust source. The velocity field here is simply }

\subsection{Slow motion and weak-field approximation, $v^\mu v_\mu \ll 1$}
Suppose that the matter frame and quasi-Newtonian frame have a relative velocity that is everywhere small: $\abs{v^\mu} \ll 1$, and that it is not rapidly changing in time either: $\abs{n^\nu \nabla_\nu (v^\mu) } / \theta \ll 1$. 
In the following, we shall keep terms up to first order in $v^\mu$ and  $n^\nu \nabla_\nu v^\mu$. 
We shall in addition assume that $\abs{v^\nu} \abs{D_\nu v^\mu}  / \theta \ll 1$, which means that scales of nonlinearity in the expansion rate of the matter frame are included {as long as $\abs{D_\nu v^\mu}  / \theta$ does not become very large. Scales where rapid (much faster than Hubble expansion) collapse of matter occurs is thus excluded.} 
{In practice, we may safely consider scales larger than galaxy clusters.}   
Under these approximations, we have 
\bea
\label{acc4}
a^\mu =  - u^\nu \nabla_\nu v^\mu - \frac{1}{3} \theta v^\mu   \,  , 
\eea  
and it follows that $\abs{a^\mu}/\theta \ll 1$ (note that spatial gradients of $v_\mu$ and $a_\mu$ are not necessarily small). We thus keep terms to first order in $a^\mu/\theta$ for consistency in the approximation. 

\vspace{6pt}
\noindent 

\noindent\fbox{%
    \parbox{0.46\textwidth}{%
\noindent 
\textbf{Weak-field interpretation.}  
The slow-motion condition implies a weak-field approximation around an FLRW space-time description in the following sense: The limit $v^\mu \! =\! 0$ together with \eqref{def:zeroshearvorticity} and \eqref{T} implies that the geometry is of the FLRW class, cf. section~3 of~\cite{Ellis:2011pi}.       
Thus, $v^\mu$ may be seen as a perturbation field relative to an FLRW solution. The condition $\abs{a^\mu}/\theta \! \ll \! 1$ implies that the logarithm of the relative time-lapse is bounded by a number much less than one for points separated by length scales $\lesssim \! 1/\theta$. Thus, structures that cause large gravitational time dilation, such as black holes, are not described in this formalism.  
    }
}

\vspace{6pt}
\noindent 
We obtain for the energy momentum tensor (\ref{T}) as projected onto the quasi-Newtonian rest frame as 
\bea
\label{Tn}
\hspace{-0.4cm} T_{ \mu \nu } = \rho \,  n_{ \mu } n_{\nu }  + \rho  ( n_{ \mu } v_{ \nu }   + n_{ \nu } v_{ \mu } )   \, , 
\eea 
where the quasi-Newtonian observers see a flux of mass $\propto \rho v_\mu$ due to their relative motion with respect to the matter frame.  
The Raychaudhuri equation for the quasi-Newtonian frame reads 
\bea
\label{raychaudhurin}    
\hspace*{-0.8cm}  n^\mu \nabla_\mu \theta  &=&  -\frac{1}{3}\,\theta^{2} - 4\pi G \rho + \Lambda  + D_\mu a^\mu   \ .    
\eea 
Consider furthermore the evolution equations for shear in the quasi-Newtonian frame, which when imposing \eqref{def:zeroshearvorticity} and \eqref{Tn} reduce to the following constraint equation 
\bea
\label{sheardot}
\hspace*{-0.8cm}  
 0 =  -   E_{\mu \nu }  + D_{\la \mu} a_{\nu \ra}  \, ,  
\eea 
where $E_{\mu \nu } \equiv n^\rho n^\sigma  C_{\rho \mu \sigma \nu} $ is the electric part of the Weyl tensor in the quasi-Newtonian frame. 
It follows from \eqref{magnetic} and \eqref{sheardot} that the space-time is conformally flat when the trace-free part of the spatial derivative of $a^\mu$ vanishes. 

We may define the Ricci curvature of the three-dimensional spatial hypersurfaces, ${}^{(3)}\! R_{\mu \nu}$.  
The associated spatial Ricci scalar, ${}^{(3)}\! R \equiv h^{\mu \nu} {}^{(3)}\! R_{\mu \nu}$, satisfies 
\bea
\label{Hamiltonianconstraint} 
{}^{(3)}\! R = 16\pi G \rho - \frac{2}{3} \theta^2 + 2 \Lambda 
\eea 
by the Hamiltonian constraint equation. 
We furthermore have the constraints 
\bea
\label{momentumconstraint}
\hspace{-0.4cm}  D_\mu \theta = 12 \pi G \rho v_\mu \, ,  \quad \;   D_\nu E_{\mu}^{ \nu } = \frac{8}{3}  \pi G ( D_\mu \rho -  \theta \rho v_\mu ) \, ,
\eea 
from which it can be derived that the vorticity in the matter frame reads 
\bea
\label{vorticityfluid}
\nabla_{  [ \mu}u_{\nu  ] }  = \frac{1}{12 \pi G \rho^2}  D_{ [ \mu } \theta D_{\nu  ]  } \rho    \, . 
\eea  
Thus, if vorticity vanishes in the matter frame, then this implies alignment of the spatial gradients of expansion and mass density. 
We notice from the first constraint in \eqref{momentumconstraint} that spatial derivatives of the expansion rate are suppressed by a factor of $v_\mu$, so the quasi-Newtonian frame is almost homogeneously expanding.

\section{Observed distances and redshifts}
\label{sec:zdist} 

{We shall now consider the distances and redshifts to sources in the matter frame as measured by observers that are themselves located in the matter frame. 
In particular, we shall investigate the anisotropies in the distance--redshift relation} and their relation to the peculiar-velocity field, $v^\mu$. 
To do this, we shall consider the (almost) kinematically-stable quasi-Newtonian frame as a reference for distances and redshifts, and we shall derive the cosmographic expression for the distance--redshift relation in this frame. We shall then relate the quasi-Newtonian distance--redshift relation to that of observers and emitters in the matter frame by their velocities. 
We refer the reader to \cite{2009GReGr..41..581E,Perlick:2010zh} for details on the calculation of distances and redshifts in the geometrical optics approximation in Lorentzian space-times, and to \cite{Heinesen:2020bej} for results on the luminosity distance cosmography. 

\subsection{Quasi-Newtonian distance--redshift relation} 
Let a beam of null-geodesic light with 4-momentum $k^\mu$ pass from a point of emission $\emi$ to a point for observation $\obs$ in this space-time. 
We shall initially imagine that the emitting and observing sources are comoving in the idealised quasi-Newtonian frame. 
We decompose the 4-momentum of the light as 
\bea
\label{kdecomp}
k^\mu = E (n^\mu - e^\mu) \, , 
\eea 
where $e^\mu$ is a spatial unit vector satisfying $n^\mu e_\nu = 0$ and $E \equiv - n^\mu k_\mu$ is the energy measured by observers comoving in the quasi-Newtonian frame. 
The redshift, $z$, of the light in the quasi-Newtonian frame is given by  
\bea
\label{def:zprime}
\hspace*{-0.2cm} \frac{ {\rm d} z}{{\rm d} \lambda} = - E_\obs (1+z)^2  \Eu \, ,   \qquad z  \equiv  \frac{E}{E_\obs} - 1  \, , 
\eea 
where $\lambda$ is as an affine parameter along the beam and satisfies $k^\mu \nabla_\mu \lambda = 1$, and where $\frac{ {\rm d} }{{\rm d} \lambda} \! \equiv \! k^\nu \nabla_\nu$ is the directional derivative along the null-geodesic.  
We have introduced the \emph{effective Hubble parameter} 
\bea
\label{def:Eu}
\hspace*{-0.2cm}  \Eu \equiv      \frac{ {\rm d} E^{-1}}{{\rm d} \lambda}    = \frac{1}{3} \theta - e^\mu a_\mu  \, , 
\eea 
which reduces to the FLRW Hubble parameter \sayy{$\dot{a}/a$} in the FLRW geometry, with $a$ being the FLRW scale factor. 
{We see that there is a sense in which the inverse photon energy function $E^{-1}$ plays the role of a generalised scale factor on the past light cone of the observer. }
{The effective Hubble parameter $\Eu$ evolves from the emitter to the observer in a way that is generally nontrivial and depends on how the wavelength of the photon stretches/contracts in response to the cosmic structure that the photon encounters. 
The evolution of $\Eu$ may be quantified in terms of an \emph{effective deceleration parameter}, which is related to the curvature of the space-time through differentiation of \eqref{def:Eu}; cf. equations~(3.9) and (4.2) in \cite{Heinesen:2020bej} for details.  
See also \cite{Heinesen:2022lqs} for a discussion on the evolution of $\Eu$ in relation to theorems that are bounding the distance-redshift relation for broad classes of universe geometries. } %The implications for the evidence of cosmic acceleration from cosmic data of the generalised response of $\Eu$ to cosmic structures are discussed in \cite{Heinesen:2022lqs}.  
We may integrate \eqref{def:zprime} to yield the following covariant decomposition of the redshift  
\bea
\label{def:zintQN}
\hspace{-0.5cm}  1   +  z 
= \frac{N_\obs}{N}  &\times& \text{exp}\left({ - \int_\lambda^{\lambda_\obs} \!\! {\rm d}\lambda' E     n^\mu \nabla_\mu \ln(N)  }\right)  \nonumber \\  
&\times&  \text{exp}\left({ \int_\lambda^{\lambda_\obs} \!\! {\rm d}\lambda' E   \! \frac{1}{3}  \theta  }\right)  . 
\eea 
The first factor is the ratio between the lapse of the observer's and emitter's proper times, and may be viewed as the gravitational redshift contribution. The second factor relates to the time evolution of the time lapse and may be viewed as an integrated Sachs--Wolfe effect. The third factor comes from the expansion of space along the light beam. 
 
We define the angular diameter distance in the quasi-Newtonian frame as $d_A \equiv \sqrt{ \delta A / \delta \Omega }$, where $\delta A$ is the cross-sectional area of the bundle of rays passing from the source and $\delta \Omega $ is the area subtended by the source when observed in the quasi-Newtonian frame. 
The luminosity distance is defined as $d_L \equiv \sqrt{L/(4\pi F)}$, where $L$ is the bolometric luminosity of light emitted from the source, and $F$ is the bolometric flux that is observed; here all quantities are again evaluated in the quasi-Newtonian frame.

Assuming that the photon number is conserved on the path from the emitter to the observer, Etherington's reciprocity theorem can be applied to arrive at the relation:  
\bea
\label{def:dL}
d_L = (1+z)^2 d_A \, . 
\eea 
{Thus, for any such null geodesic congruence ending in a vertex at an observer in a Lorentzian geometry, we may use this relation and the geometrical definition of the angular diameter distance to obtain the Luminosity distance. See \cite{Perlick:2010zh} for a  comprehensive modern reference on geometrical optics in Lorentzian geometries, and Sachs optical equations for angular diameter distance.   }

\subsection{Luminosity distance cosmography} 
Assuming analyticity, we can write $d_L$ in terms of its Taylor series expansion around the observer, yielding the luminosity distance cosmography in the quasi-Newtonian frame{; see \cite{Heinesen:2020bej} for details:}  
\bea
\label{dLexpand3} 
\hspace*{-0.5cm} d_L  &=& d_L^{(1)} z   + d_L^{(2)} z^2 +  d_L^{(3)} z^3 + \Delta d_L^{(3)}(z)  \ , \nonumber \\ 
\hspace*{-0.5cm}  d_L^{(1)} &\equiv& \frac{1}{\Eu_\obs} \, , \qquad d_L^{(2)} \equiv   \frac{1}{ \Eu_\obs}  \left(  1 +  \frac{1}{2} \frac{1}{E} \frac{     \frac{ {\rm d} \Eu}{{\rm d} \lambda}    }{\Eu^2}  \Bigr\rvert_{\obs} \right) \ , \nonumber \\ 
\hspace*{-0.5cm}  d_L^{(3)} &\equiv&  \frac{1}{ \Eu_\obs}  \left( \!  \frac{1}{2}  \frac{1}{E^2} \frac{  \left(    \frac{  {\rm d} \Eu}{{\rm d} \lambda} \right)^2    }{\Eu^4}  \Bigr\rvert_{\obs} \right. \nonumber  \\ 
&& \left. -  \frac{1}{6}  \frac{1}{E^2} \frac{      \frac{  {\rm d^2} \Eu}{{\rm d} \lambda^2}    }{\Eu^3}  \Bigr\rvert_{\obs}  -  \frac{1}{12} \frac{1}{E^2} \frac{k^{\mu}k^\nu R_{\mu \nu} }{\Eu^2}  \Bigr\rvert_{\obs}   \right)     \,  , 
\eea  
where $\Delta d_L^{(3)}(z)$ is the remainder term of the third order expansion. 
{We go to third order, since this is sufficient for analysing large scale bulk~flows in the low-redshift regime\footnote{{Note that convergence properties and level of approximation at a given redshift must in principle be assessed independently for any model-geometry. See appendix~A of~\cite{Macpherson:2021gbh} for a discussion of convergence properties in the Einstein~Toolkit cosmology simulations.}}, $z \ll 1$.}  
The coefficient $d_L^{(1)}$ is anisotropic with the dipolar term $e^\mu a_\mu \rvert_{\obs}$, as can be seen from \eqref{def:Eu}. 
We consider scales of observation smaller than $\theta^{-1}_\obs$, and thus the leading order anisotropic terms in $d_L^{(2)}, d_L^{(3)}$, etc., will be those with a maximum number of spatial derivatives of $a^\mu$. Keeping those leading order terms only for each coefficient, together with the nonperturbative contributions from $\theta$ and ${}^{(3)}\! R$, we have  
\bea
\label{dLcoef} 
\hspace*{-0.2cm}  \Eu_\obs  d_L^{(2)}  &\approx&  \frac{ 1 -  q_m  }{ 2 } +  \frac{  e^\mu e^\nu D_{\la \mu} a_{\mu \ra}  }{ 2 \left( \frac{\theta}{3}\right)^2 }  \Bigr\rvert_{\obs}   \ , \nonumber \\ %\frac{1}{3}D_\mu a^\mu + 
\hspace*{-0.2cm}  \Eu_\obs d_L^{(3)}  &\approx&  \frac{  -1 + 3 q_m^2 +  q_m + \mathcal{R}_m - j_m    }{ 6 }   \nonumber     \\ 
\hspace*{-0.1cm}    &+& \frac{   \frac{19}{45} e^\mu D_\mu D_\nu a^\nu +   e^\mu e^\nu e^\rho D_{\la \mu} D_\nu a_{\rho \ra }   }{6 \left( \frac{\theta}{3}\right)^3} \Bigr\rvert_{\obs}        \,  , 
\eea 
where the monopolar parameters 
\bea
\label{params} 
q_m  &\equiv& - 1 - 3 \frac{n^\mu \nabla_\mu \theta + D_\mu a^\mu}{\theta^2}  \Bigr\rvert_{\obs}   \, , \quad  \mathcal{R}_m \equiv - \frac{1}{6} \frac{ {}^{(3)}\! R }{(\frac{\theta}{3})^2 }  \Bigr\rvert_{\obs}   \, , \nonumber \\ 
 j_m  &\equiv&   1 + 9 \frac{n^\nu \nabla_\nu (n^\mu \nabla_\mu \theta) + \theta n^\mu \nabla_\mu \theta }{\theta^3}  \Bigr\rvert_{\obs}   \, , 
\eea 
determine the isotropic parts of the luminosity distance function. 
{The isotropic contributions are on a form that is similar to the FLRW deceleration, curvature, and jerk parameters when $\theta/3$ is thought of as a Hubble parameter and when ${}^{(3)}\! R$ is thought of as the FLRW spatial curvature. In general, however, $\theta$ and ${}^{(3)}\! R$ may be inhomogeneous functions (so that they vary across observers). 
The spatial divergence of $a^\mu$ also enters in the expression for $q_m$ and can be seen as an isotropic tidal contribution.  } 
{The remaining anisotropic contributions in \eqref{dLcoef} come from spatial gradients in $a^\mu$, which may be interpreted as tidal effects from the inhomogeneity of the gravitational field strength. The anisotropic terms are discussed individually in some detail in the section with the discussion of the main results below.}
In calculating the coefficient $d_L^{(3)}$ in \eqref{dLcoef} we have used the identity $3 h^{\alpha \beta} D_{\alpha} D_{\la \beta} a_{\mu \ra} = 3 {}^{(3)}\! R^\alpha_{\, \mu} a_\alpha + 2D_\mu (D_\alpha a^\alpha)$.  
\subsection{The distance--redshift relation in the matter frame}
We consider the generic transformation of redshift and angular diameter distance under a small boost of the emitter and observer.  
We let this boost be the one from the quasi-Newtonian frame to the matter frame\footnote{Gravitational potentials within the virialized neighbourhood of the observer/emitters will generate additional contributions to their velocities, which are not accounted for in the large-scale matter model. For the purpose of estimating large-scale matter flow, these may be considered noise terms.}, such that 
\bea
\label{def:zhat2}
\hspace*{-0.45cm} 1 + \hat{z} = (1+z) (1 + e^\mu v_\mu \rvert_{\emi}  \! -\!  e^\mu v_\mu \rvert_{\obs} \!) \,  , \; \;  
\eea 
and 
\bea
\label{def:dAhat}
\hat{d}_A   =  (1  +  e^\mu v_\mu \rvert_{\obs} \!)  \, d_A \, , 
\eea   
where the hat indicates that the emitter and the observer are located in the matter frame, and where the expressions are linear in the velocity field. 
The observed luminosity distance reads 
\bea
\label{def:dLhat}
\hat{d}_L = (1+\hat{z})^2 \hat{d}_A  = (1 + 2e^\mu v_\mu \rvert_{\emi}  \! -\!   e^\mu v_\mu \rvert_{\obs} \!) d_L \, . 
\eea 
We may consider the situation where the observer's motion has already been accounted for, by appropriate transformations of the observables, that will then read: $1 + \hat{z} = (1+z) (1 + e^\mu v_\mu \rvert_{\emi}  \!)$ and $\hat{d}_L = (1 + 2e^\mu v_\mu \rvert_{\emi}  \!) d_L$. 
We substitute these transformations into $(\theta_\obs/3) \times$\eqref{dLexpand3} to obtain the following expression 
\bea
\label{dLpeculiar} 
\hspace*{-0.55cm}  &&  \frac{1}{3} \theta_\obs \hat{d}_L   +  e^\mu v_\mu \rvert_{\emi}(1 -  \hat{z}  + 2 \Eu_\obs d_L^{(2)}  \hat{z}    )          \nonumber  \\
  \hspace*{-.55cm} &&  =  \hat{z} + \frac{e^\mu a_\mu \rvert_{\obs} }{ \frac{1}{3} \theta_\obs} \hat{z}    + \Eu_\obs   \left( d_L^{(2)} \hat{z}^2 +  d_L^{(3)} \hat{z}^3 + \Delta d_L^{(3)}(\hat{z})  \right)  , %\Eu_\obs  d_L(\hat{z})  =
\eea 
where we have used the definition \eqref{def:Eu}. 
We have kept terms of order $\hat{z} e^\mu v_\mu \rvert_{\emi} $, but we have neglected terms $\propto \hat{z}^n e^\mu v_\mu \rvert_{\emi}$ with $n \geq 2$. 

\vspace{6pt}
\noindent 

\noindent\fbox{%
    \parbox{0.46\textwidth}{%
\noindent 
\textbf{The exact FLRW case.}  
To recover the results for an FLRW space-time, we formally set $a^\mu = 0$ in the above equations, and view $v^\mu \rvert_{\emi}$ as small velocities relative to the comoving FLRW frame. Then $\theta_\obs/3$ reduces to the Hubble parameter, and $d_L^{(n)}$ reduce to their usual FLRW expressions given in \cite{Visser:2004bf}. 
In this limit, our result in \eqref{dLpeculiar} agrees with the formula in equation (15) of~\cite{Hui:2005nm} for motions relative to an exact FLRW frame, which can be seen by performing a Taylor series expansion in redshift of their formula and keeping cross terms with the velocity field up to order $e^\mu v_\mu \rvert_{\emi}  \hat{z}$. 
    }%
}

\vspace{6pt}
\noindent 
The formula \eqref{dLpeculiar} may be considered final for isolating the velocity component $e^\mu v_\mu \rvert_{\emi}$. 
There is, however, the subtlety that the emitter's position, $\emi$, is at the past light cone. 
To use \eqref{dLpeculiar} to estimate bulk flows of a volume \emph{at a fixed cosmological time}, we must account for the depth of the survey in time. 
We have to first order in elapsed time, $\tau_\obs - \tau_\emi$, in the frame of the source that 
\bea
\label{vevol} 
\hspace*{-0.5cm} &&  e^\mu v_\mu \rvert_{\emi} = e^\mu v_\mu \rvert_{p^\emi_\obs}   +   u^\nu \nabla_\nu ( e^\mu v_\mu ) \rvert_{p^\emi_\obs} (\tau_\emi - \tau_\obs) \, , 
\eea 
where the point $p^\emi_\obs$ is the space-time point of the source \sayy{today}. 
Rewriting \eqref{vevol} in terms of the distance in redshift to the emitter, we have 
\bea
\label{vevolz} 
\hspace*{-0.5cm} &&  e^\mu v_\mu \rvert_{\emi} = e^\mu v_\mu \rvert_{p^\emi_\obs}   -   \frac{ u^\nu \nabla_\nu (e^\mu v_\mu)  \rvert_{p^\emi_\obs}  }{ \frac{1}{3} \theta_\obs  }  \hat{z} \, , 
\eea
where we have used the first constraint in \eqref{momentumconstraint} to replace $\theta  \rvert_{p^\emi_\obs}$ with $\theta_\obs$ to zeroth order in $\abs{v^\mu}$. 
Decomposing $u^\nu \nabla_\nu (e^\mu v_\mu)  \rvert_{p^\emi_\obs} =  v_\mu u^\nu \nabla_\nu  e^\mu   \rvert_{p^\emi_\obs}  +  e^\mu  u^\nu \nabla_\nu v_\mu   \rvert_{p^\emi_\obs}$, the first term arises from the drift in the direction of emission of photons from the source while the second term is due to the evolution of the velocity field. 
The first term is zero in linearised perturbation theory around an FLRW metric but may be important on scales of nonlinear structures \cite{2010PhRvD..81d3522Q,Krasinski:2010rc}.

\vspace{6pt}
\noindent 
\noindent\fbox{%
    \parbox{0.46\textwidth}{%
\noindent 
\textbf{Velocity evolution in perturbation theory.}  
In standard linearised perturbation theory, the velocity evolution reads $u^\nu \nabla_\nu v_\mu \propto v_\mu$, with a magnitude and sign that can be calculated via the velocity growth function \cite{1980lssu.book.....P,1991MNRAS.251..128L}. 
For the spatially flat $\Lambda$CDM model with $\Omega_{\text{matter}} \! = \! 0.3$ and $\Omega_{\Lambda} \! = \! 0.7$, it happens that $u^\nu \nabla_\nu v_\mu \rvert_{p^\emi_\obs} = - 0.07  (\dot{a}/a)  v_\mu \rvert_{p^\emi_\obs}$; cf. equations~(6) and (10) in~\cite{1991MNRAS.251..128L}. %The term is thus subdominant in the standard model. 
For comparison, $\Omega_{\text{matter}} \! = \! 1$ gives $u^\nu \nabla_\nu v_\mu \rvert_{p^\emi_\obs} = 0.5   (\dot{a}/a)  v_\mu \rvert_{p^\emi_\obs}$, whereas $\Omega_{\Lambda} \! =\! 1$ yields $u^\nu \nabla_\nu v_\mu \rvert_{p^\emi_\obs} = - 1.0 (\dot{a}/a)  v_\mu \rvert_{p^\emi_\obs}$. 
    }%
}

\vspace{6pt}
\noindent 
Inserting \eqref{vevolz} in \eqref{dLpeculiar}, we have 
\bea 
\label{dLpeculiarobs} 
\hspace*{-0.4cm}  &&  \frac{1}{3} \theta_\obs \hat{d}_L   +  e^\mu v_\mu \rvert_{p^\emi_\obs} \! \! \left(\! 1 -  \hat{z}  + 2 \Eu_\obs d_L^{(2)}  \hat{z}  \!  \right)      -   \frac{ u^\nu \nabla_\nu (e^\mu v_\mu)  \rvert_{p^\emi_\obs}  }{\frac{1}{3} \theta_\obs }  \hat{z}     \nonumber  \\
 &&    =   \hat{z}  +   \frac{e^\mu a_\mu \rvert_{\obs} }{ \frac{1}{3} \theta_\obs} \hat{z}  + \Eu_\obs   \left( d_L^{(2)} \hat{z}^2 +  d_L^{(3)} \hat{z}^3 + \Delta d_L^{(3)}(\hat{z})  \right)  . %
\eea 
This relation is the main result of this paper, and expresses the peculiar-motion of the astrophysical source evaluated at the present epoch in terms of its measured distance and redshift.

\section{Discussion of the main result} 
\label{sec:discussion} 
The relation \eqref{dLpeculiarobs} provides a cosmographic expression for peculiar-velocities given our weak-field description.  
Loosely speaking, it applies to scales where gravitational waves and frame dragging can be ignored, and where gravitational time dilation is perturbative. 
It \emph{does} apply to scenarios with nonlinear contrast in density and in spatial curvature. 
Thus, various systems that are usually considered nonperturbative in terms of curvature are included in this weak-field description, for instance, certain Lema\^itre-Tolman-Bondi structures \cite{VanAcoleyen:2008cy,2009JCAP...09..022E}. 
Here we discuss the terms appearing in \eqref{dLpeculiarobs} and their significance for recent analyses of bulk motions in the literature. 																							

\subsection{The individual terms in peculiar-velocity formula}
The anisotropic terms on the left-hand side of the equality in \eqref{dLpeculiarobs} arise due to pure motion effects. 
The anisotropic terms on the right-hand side may be thought of as intrinsic, i.e., they are there even when the observer and emitter 4-velocities coincide with that of the quasi-Newtonian reference frame, and their dominant contributions come from the gravitational redshift of sources; cf. equation~\eqref{def:zintQN}.  

It is tempting to set $a^\mu$ to zero in the neighbourhood of observation in order to avoid dealing with these terms. However, this \emph{cannot} be done in a consistent treatment of the bulk motions, as the same gravitational structures that cause the bulk motions are causing time dilation in the quasi-Newtonian frame. Indeed, we can expect $3\abs{a^\mu}/\theta \sim \abs{v^\mu}$ from \eqref{acc4}. 
We shall here give a brief interpretation of the anisotropic terms on the right-hand side of \eqref{dLpeculiarobs}.

\vspace{1pt}

\noindent 
\textbf{The term $3 e^\mu a_\mu \rvert_{\obs}  \hat{z} / \theta_\obs$}  describes the dipole in the gravitational redshift of sources as is evident from the equation for redshift~\eqref{def:zintQN} and the definition of $a_\mu$ in terms of the lapse function. 
{The gravitational redshift is caused by the same gravitational potentials that cause the flow of matter, and the dipole in the gravitational redshift may thus be expected to show alignment with the peculiar flow. }
Indeed, in linearised perturbation theory around an FLRW background, the following proportionality relation holds: $3 a_\mu / \theta \propto v_\mu$, with a proportionality factor which is $\approx - 1$ at late cosmic times in the spatially flat $\Lambda$CDM model with $\Omega_{\text{matter}} \! = \! 0.3$; cf. \cite{1991MNRAS.251..128L}. 
%The dipole term $3 e^\mu a_\mu \rvert_{\obs}  \hat{z} / \theta_\obs$ can thus be expected to be of similar order and direction as $- e^\mu v_\mu \rvert_{\obs} \hat{z}$.  
%As opposed to the terms that are proportional to $e^\mu v_\mu \rvert_{\emi}$, there
We note that there is no statistical cancellation effect for this term, as the term is evaluated at the observer position.   

\vspace{1pt}

\noindent 
\textbf{The terms $\Eu_\obs d_L^{(2)} \hat{z}^2$ and $\Eu_\obs d_L^{(3)} \hat{z}^3$}  
incorporate higher-order {derivatives of the gravitational redshift}. 
The anisotropic signature of the term $\Eu_\obs d_L^{(2)} \hat{z}^2$ is a quadrupole {(see \eqref{dLcoef}). By} \eqref{sheardot} we see that it is probing the electric Weyl tensor in the quasi-Newtonian frame{. The electric Weyl tensor is analogous to  \emph{minus} the Newtonian tidal tensor and it is thus describing deviations of the gravitational field strength with change of position.}  
The anisotropic signature of $\Eu_\obs d_L^{(3)} \hat{z}^3$ is given by a combination of a dipole and an octupole. From the second relation in~\eqref{momentumconstraint}, we derive that {the dipole} contribution is along $v_\mu$ and $D_\mu \rho$. Vanishing vorticity in the matter frame implies that $D_\mu \rho \propto v_\mu$ by~\eqref{vorticityfluid}. 
{Thus, for the vanishing vorticity case,} it is expected that the dipolar contribution coming from the term $\Eu_\obs d_L^{(3)}$ is proportional to $e^\mu v_\mu \rvert_{\obs}$.  

\vspace{1pt}

\noindent 
{\textbf{Comparison of the size of the anisotropies.}}
Evaluating the derivatives of $a_\mu$ over a scale comparable to the observed redshift $\hat{z}$, we have $\abs{e^\mu e^\nu D_\mu a_\nu}  \rvert_{\obs} \! \sim \! \abs{a_\nu} \frac{\theta}{3} \rvert_{\obs} / \hat{z}$; $\abs{e^\rho e^\mu e^\nu D_\rho D_\mu a_\nu}  \rvert_{\obs} \sim \abs{a_\nu} \left(\frac{\theta}{3} \right)^2 \rvert_{\obs} / \hat{z}^2$. Thus, 
the terms $\Eu_\obs d_L^{(2)} \hat{z}^2$ and $\Eu_\obs d_L^{(3)} \hat{z}^3$ are expected to be of roughly the same order as the first order contribution $3 e^\mu a_\mu \rvert_{\obs}  \hat{z} / \theta_\obs$. We, however, note the factor of $6$ in the denominator of $d_L^{(3)}$ in \eqref{dLcoef} that will suppress this term. 
{At higher order, the terms $\Eu_\obs d_L^{(n)} \hat{z}^n$ will become increasingly suppressed by the factorials when the series is convergent.}  

%, and their contribution may be evaluated for specific space-time models. 

\vspace{1pt}
\noindent

\subsection{Comments on recent empirical analyses}

Many bulk flow estimators in the literature, e.g., those in recent analyses \cite{Nusser:2011sd,Watkins:2023rll}, are considering the lowest-order anisotropic term $e^\mu v_\mu \rvert_{p^\emi_\obs}$ in \eqref{dLpeculiarobs} while neglecting the higher-order contributions. Other studies again\footnote{In appendix~A~of~\cite{Carreres:2023nmf}, the impact of peculiar-velocity estimates from neglecting various terms of order $v \times z$ in equation (15) of~\cite{Hui:2005nm}. However, this equation itself is only exact in the idealised case of peculiar-motions relative to an exact FLRW frame, and is thus missing the additional relativistic terms of order $v \times z$ as discussed in this paper.}, e.g., \cite{Watkins:2014oia,Johnson:2014kaa,Howlett:2017asw,Carreres:2023nmf}, are including some of the terms of order $v \times z$, but are neglecting others. 
In both cases, this is generally expected to lead to theoretical biases of the estimated bulk velocity of order $v \times z$.  
Such correction biases, even if subdominant when $z \lesssim 0.1$, may significantly alter the error analysis and any quoted significance levels. 

Perhaps most importantly, the discussed theoretical biases could alter the observed \emph{trend} in bulk flows as deduced in some recent analyses \cite{Peery:2018wie,Watkins:2023rll}. 
From an order of magnitude estimate within the spatially flat $\Lambda$CDM model, with $\Omega_{\text{matter}} = 0.3$, we have that $2 \Eu_\obs d_L^{(2)} - 1 \sim 0.55$, and thus $e^\mu v_\mu \rvert_{p^\emi_\obs} (-  \hat{z}  + 2 \Eu_\obs d_L^{(2)}  \hat{z}) \approx 0.55 \times  \hat{z} e^\mu v_\mu \rvert_{p^\emi_\obs}$. 
The correction term coming from the velocity evolution is subdominant for $\Omega_{\text{matter}} = 0.3$, as discussed in the box above \eqref{dLpeculiarobs}, and we set $u^\nu \nabla_\nu v^\mu \rvert_{p^\emi_\obs} \approx 0$. 
Using this approximation, we have $3 e^\mu a_\mu \rvert_{\obs}  \hat{z} / \theta_\obs \approx - e^\mu v_\mu \rvert_{\obs} \hat{z}$.  
Putting all of the approximations together in \eqref{dLpeculiarobs} gives the $\Lambda$CDM estimate: 
\bea 
\label{dLpeculiarobsFLRW} 
\hspace*{-0.6cm}  &&     e^\mu v_\mu \rvert_{p^\emi_\obs} \!\! \left(1 + 0.55  \hat{z}    \right)  + e^\mu v_\mu \rvert_{\obs}  \hat{z} =     \Eu_\obs d_L(\hat{z})  -  \frac{1}{3} \theta_\obs \hat{d}_L    \, ,  \nonumber \\ 
&& \Eu_\obs d_L(\hat{z}) \equiv  \hat{z} +\Eu_\obs   \left( d_L^{(2)} \hat{z}^2 +  d_L^{(3)} \hat{z}^3 + \Delta d_L^{(3)}(\hat{z})  \right) . 
\eea 
If neglecting anisotropies in the higher-order coefficients $\Eu_\obs d_L^{(n)}$, $n \geq 2$, we have that $\Eu_\obs d_L(\hat{z})$ is isotropic, and the residual on the right-hand side of the equality would be what is typically used in lowest-order estimates of peculiar-velocities. 
At the left-hand side of the equality, the correcting factor $\left(1 + 0.55  \hat{z} \right)$ amplifies the true velocity amplitude. 
The term $e^\mu v_\mu \rvert_{\obs}  \hat{z}$ in addition amplifies the amplitude in cases where $e^\mu v_\mu \rvert_{p^\emi_\obs}$ and $e^\mu v_\mu \rvert_{\obs}$ are of the same sign. 
{The formula in \eqref{dLpeculiarobsFLRW} is local and in practice applies to the individual galaxies of a given survey. However, averaging over the individual observed velocities $e^\mu v_\mu \rvert_{p^\emi_\obs}$, and assuming that these are uncorrelated with the observed redshifts $\hat{z}$ at lowest order, the formula applies to the bulk flow, $\overline{e^\mu v_\mu \rvert_{p^\emi_\obs}}$, and the mean redshift, $\overline{\hat{z}}$, with 
\bea 
\label{dLpeculiarobsFLRWav} 
\hspace*{-0.6cm} \overline{e^\mu v_\mu \rvert_{p^\emi_\obs}} \!\! \left(1 + 0.55  \overline{\hat{z}}    \right)  + e^\mu v_\mu \rvert_{\obs}  \overline{\hat{z}} =     \overline{\Eu_\obs d_L(\hat{z})  -  \frac{1}{3} \theta_\obs \hat{d}_L} , 
\eea
where the overbar indicates the average by the galaxy window function employed in the analysis.
}

In the case where the amplitude of the local flow\footnote{{The local flow $v_\mu \rvert_{\obs}$ does not generally coincide with the bulk flow, as the latter is rather obtained by a (weighted) mean of $v_\mu \rvert_{p^\emi_\obs}$. However, when the observational scales become as small as the fluid elements of the matter model, the emitters and observer can essentially be treated as belonging to the same fluid parcel, and we expect the value of the bulk flow to approach that of the local flow.}} $v_\mu \rvert_{\obs}$ is of order of the mean estimated emitter flows $v_\mu \rvert_{p^\emi_\obs}$ and in the same direction\footnote{{This case is motivated by the observed bulk flows in~\cite{Nusser:2011sd,Watkins:2023rll}, where the direction of the bulk flow remains approximately stable across the investigated scales 40--200 Mpc/$h$ (and probably stable across even broader ranges, cf. discussion on the flow of the local sheet in the below discussion), and where the amplitude of the flow is of the order of a few hundred km/s across the same scales.}}, {we have that the left-hand side of \eqref{dLpeculiarobsFLRWav} yields $\sim \overline{e^\mu v_\mu \rvert_{p^\emi_\obs}} \!\! \left(1 + 1.55  \overline{\hat{z}}    \right)$, thus giving rise to} an amplifying factor of {$\sim  \left(1 + 1.55  \overline{\hat{z}}    \right)$ for the mean estimated $e^\mu v_\mu \rvert_{p^\emi_\obs}$}.  
{Within this order-of-magnitude analysis, the overestimate of the bulk velocity amplitude at $\hat{z} = 0.1$ (corresponding to $300$Mpc/$h$) would be $\sim 15\%$.}

Indeed, in \cite{Watkins:2023rll} the estimate of the bulk flow at $200$ Mpc/$h$ is $420$ km/s in the direction $(l = 298, b=-8)$, with a direction of the flow that is rather stable across the investigated redshift range.  
This inferred large-scale flow is notably similar to the estimated flow of the local sheet of galaxies: In the rest frame of the local sheet, the dipole in the cosmic microwave background corresponds to the velocity $631$~km/s in the direction $(l = 270, b=27)$, which, once corrected for close by \sayy{attractors} and \sayy{repellors} within a radius of $z = 0.01$, yields the residual velocity of $455$~km/s with direction $(l = 299,  b = 15)$ \cite{Tully:2007tp}. 
We take the latter inferred velocity as an estimate for $v_\mu \rvert_{\obs}$, the above estimate roughly applies, and there should be a $\sim 10\%$ overestimate of the bulk flow at the scale $200$ Mpc/$h$ (corresponding to $z=0.06$). 
This does not seem quite sufficient in explaining the entire upgoing trend line from 100 Mpc/$h$ to 200 Mpc/$h$, even if the estimate is conservative. 
In any case, we expect the size of the correction terms to be enough to alter the trend of the estimated bulk flow and the error analysis in \cite{Watkins:2023rll}, but they would not be enough to make the results agree with the $\Lambda$CDM expected bulk flow amplitude or its decay $\propto 1/(\text{distance})$. In that sense, the main conclusions of \cite{Watkins:2023rll} are robust.

\section{Conclusion} 
\label{sec:conclusion} 
The main result of this paper is the formula~\eqref{dLpeculiarobs}, valid for inferring large-scale peculiar-motions from cosmological datasets within the cosmographic regime $z \lesssim 1$. 
When the aim is precise (mean) velocity, $v$, estimates at the level of $v \times z$ precision, all of the terms in \eqref{dLpeculiarobs} must \emph{a priori} be included in the analysis. 
These terms include both classical peculiar-motion terms, but also relativistic effects in the light propagation.

The leading order relativistic correction terms are due to gravitational time dilation\footnote{As detailed in \cite{Pyne:2003bn,Hui:2005nm}, the integrated Sachs--Wolfe effect and gravitational lensing effects also generally impact the distance--redshift curve, and they implicitly appear on the right-hand side of \eqref{dLpeculiarobs}. These latter effects are, however, subdominant in the cosmographic regime that we are considering.} and the fact that objects are observed on the past light cone. 
These effects have been analysed in detail in \cite{Hui:2005nm,Kaiser:2014jca} within linearised FLRW perturbation theory. 

Here we have provided the first cosmographic expression where all anisotropic terms up to order $v \times z$ are included consistently, and this is done within a fairly general weak-field formalism. 
Structures of arbitrary size and nonperturbative structures (in terms of density and spatial curvature contrasts) can be described within our formalism, which makes it suitable for interpreting bulk motions outside of the strictly linearised perturbative regime of an FLRW metric.  

Theoretical biases in neglecting (one or more of) the relativistic correction terms are already surpassing the estimated measurement biases in the most recent analyses of bulk flows \cite{Watkins:2023rll}. 
The inclusion of these correction terms will be important for future analyses that will constrain the bulk flow toward $z \sim 1$  \cite{Nusser:2011sd,Mathews:2014fma}. {It should be noted that the relativistic terms will be competing with the statistical measurement errors, which are currently large when going beyond $z \lesssim 0.1$. Irrespective of future efforts in reducing the error bars, the relativistic correction terms should as a minimum be included in the error analysis in order to achieve reliable significance levels for the bulk flow estimates.}  

The formula~\eqref{dLpeculiarobs} can be used to constrain bulk flows to a high precision when the quality of the data allows it, either by fixing a model beforehand, or by determining the coefficients of the cosmography in a combined fit to the data.

\vspace{6pt} 
%%%%%%%%%%%%%%%%%%%%%%%%%%%%%%%%%%%%%%%%%%%%%%%%%%%%%%%%%%%%%%%%%%%
\begin{acknowledgments}
%%%%%%%%%%%%%%%%%%%%%%%%%%%%%%%%%%%%%%%%%%%%%%%%%%%%%%%%%%%%%%%%%%% 
I am thankful to Hamed Barzegar, Thomas Buchert, {Hume~A.~Feldman} and Sofie Marie Koksbang for their careful reading of the paper and for their detailed comments which helped improve it.  
This work is part of a project that has received funding from the European Research Council (ERC) under the European Union's Horizon 2020 research and innovation programme (grant agreement ERC advanced Grant No. 740021--ARTHUS, PI: Thomas Buchert). 
{AH is funded by the Carlsberg Foundation.}
\end{acknowledgments}

\bibliographystyle{mnras_unsrt}
\bibliography{refs}

\end{document}